\documentclass[]{aa}

\usepackage{multirow}
\usepackage{graphicx}
\usepackage{txfonts}
\usepackage[]{hyperref}

\begin{document}

\newcommand{\lya}{Lyman-$\alpha$}
\newcommand{\hst}{\emph{HST}}

\title{\emph{HST} PanCET program: Non-detection of atmospheric escape in the warm Saturn-sized planet WASP-29~b}

\titlerunning{Far-ultraviolet transit spectroscopy of WASP-29~b}
\authorrunning{L. A. dos Santos et al.}

\author{L.~A.~dos~Santos
\inst{1}
\and
V.~Bourrier
\inst{1}
\and
D.~Ehrenreich
\inst{1}
\and
J.~Sanz-Forcada
\inst{2}
\and
M.~L\'opez-Morales
\inst{3}
\and
D.~K.~Sing
\inst{4, 5}
\and
A.~Garc\'ia~Mu\~noz
\inst{6}
\and
G.~W.~Henry
\inst{7}
\and
P.~Lavvas
\inst{8}
\and
A.~Lecavelier~des~Etangs
\inst{9}
\and
T.~Mikal-Evans
\inst{10}
\and
A.~Vidal-Madjar
\inst{9}
\and
H.~R.~Wakeford
\inst{11}
}

\institute{Observatoire astronomique de l’Université de Genève, Chemin Pegasi 51b, 1290 Versoix, Switzerland\\
\email{Leonardo.dosSantos@unige.ch}
\and 
Centro de Astrobiolog\'{i}a (CSIC-INTA), E-28692 Villanueva de la Ca\~nada, Madrid, Spain
\and 
Center for Astrophysics | Harvard \& Smithsonian, 60 Garden Street, Cambridge, MA 02138, USA
\and 
Department of Earth \& Planetary Sciences, Johns Hopkins University, Baltimore, MD, USA
\and 
Department of Physics \& Astronomy, Johns Hopkins University, Baltimore, MD, USA
\and 
AIM, CEA, CNRS, Université Paris-Saclay, Université de Paris, F-91191 Gif-sur-Yvette, France
\and 
Tennessee State University, Center of Excellence in Information Systems, Nashville, TN  37203, USA
\and 
Groupe de Spectrom\'etrie Mol\'eculaire et Atmosph\'erique, Universit\'e de Reims, Champagne-Ardenne, CNRS UMR F-7331, France
\and 
Institut d'Astrophysique de Paris, CNRS, UMR 7095 \& Sorbonne Universités UPMC Paris 6, 98 bis bd Arago, 75014 Paris, France
\and 
Kavli Institute for Astrophysics and Space Research, Massachusetts Institute of Technology, Cambridge, MA, USA
\and 
School of Physics, University of Bristol, HH Wills Physics Laboratory, Tyndall Avenue, Bristol BS8 1TL, UK
}

\date{Received 03 February 2021; accepted 29 March 2021}
 
\abstract{Short-period gas giant exoplanets are susceptible to intense atmospheric escape due to their large scale heights and strong high-energy irradiation. This process is thought to occur ubiquitously, but to date we have only detected direct evidence of atmospheric escape in hot Jupiters and warm Neptunes. The latter planets are particularly more sensitive to escape-driven evolution as a result of their lower gravities with respect to Jupiter-sized planets. But the paucity of cases for intermediate, Saturn-sized exoplanets at varying levels of irradiation precludes a detailed understanding of the underlying physics in atmospheric escape of hot gas giants. Aiming to address this issue, our objectives here are to assess the high-energy environment of the warm ($T_\mathrm{eq} = 970$ K) Saturn WASP-29~b and search for signatures of atmospheric escape. We used far-ultraviolet (FUV) observations from the \emph{Hubble Space Telescope} to analyze the flux time series of \ion{H}{I}, \ion{C}{II}, \ion{Si}{III}, \ion{Si}{IV}, and \ion{N}{V} during the transit of WASP-29~b. At 88 pc, a large portion of the Lyman-$\alpha$ core of the K4V-type host WASP-29 is attenuated by interstellar medium absorption, limiting our ability to probe the escape of H at velocities between [-84, +35] km~s$^{-1}$. At 3$\sigma$ confidence, we rule out any in-transit absorption of \ion{H}{I} larger than 92\% in the Lyman-$\alpha$ blue wing and 19\% in the red wing. We found an in-transit flux decrease of $39\%^{+12\%}_{-11\%}$ in the ground-state \ion{C}{II} emission line at 1334.5 \AA. But due to this signal being significantly present in only one visit, it is difficult to attribute a planetary or stellar origin to the ground-state \ion{C}{II} signal. We place 3$\sigma$ absorption upper limits of 40\%, 49\%, and 24\% on \ion{Si}{III}, \ion{Si}{IV}, and for excited-state \ion{C}{II} at 1335.7 \AA, respectively. Low activity levels and the faint X-ray luminosity suggest that WASP-29 is an old, inactive star. Nonetheless, an energy-limited approximation combined with the reconstructed EUV spectrum of the host suggests that the planet is losing its atmosphere at a relatively large rate of $4 \times 10^9$~g~s$^{-1}$. The non-detection at Lyman-$\alpha$ could be partly explained by a low fraction of escaping neutral hydrogen, or by the state of fast radiative blow-out we infer from the reconstructed Lyman-$\alpha$ line.}


\maketitle

\section{Introduction}

The large population of transiting exoplanets yields the best opportunity to perform demographic studies aiming to characterize and understand their atmospheres. One of the main techniques to do so is transmission spectroscopy, which consists in measuring wavelength-dependent changes in the stellar spectrum when a planet transits its host star. The largest atmospheric signals of exoplanets detected to date have been obtained with transmission spectroscopy in far- and near-ultraviolet (FUV, NUV). This wavelength range traces the outermost parts of the atmospheres of transiting planets, where the lightest particles are susceptible to atmospheric escape. This has been observed in planets with H- and He-dominated atmospheres, such as the warm Neptunes GJ~436~b \citep{2014ApJ...786..132K, 2015Natur.522..459E, 2016A&A...591A.121B, 2017A&A...605L...7L, 2019A&A...629A..47D}, GJ~3470~b \citep{2018A&A...620A.147B}, the hot-Jupiters HD~209458~b \citep{2003Natur.422..143V, 2004ApJ...604L..69V, 2008A&A...483..933E, 2013A&A...560A..54V}, HD~189733~b \citep{2010A&A...514A..72L, 2012A&A...543L...4L, 2013A&A...551A..63B}, WASP-12~b \citep{2010ApJ...714L.222F} and WASP-121~b \citep{2019AJ....158...91S}, and tentatively for the temperate sub-Neptune K2-18~b \citep{2020A&A...634L...4D}.

\begin{figure}
\includegraphics[width=\hsize]{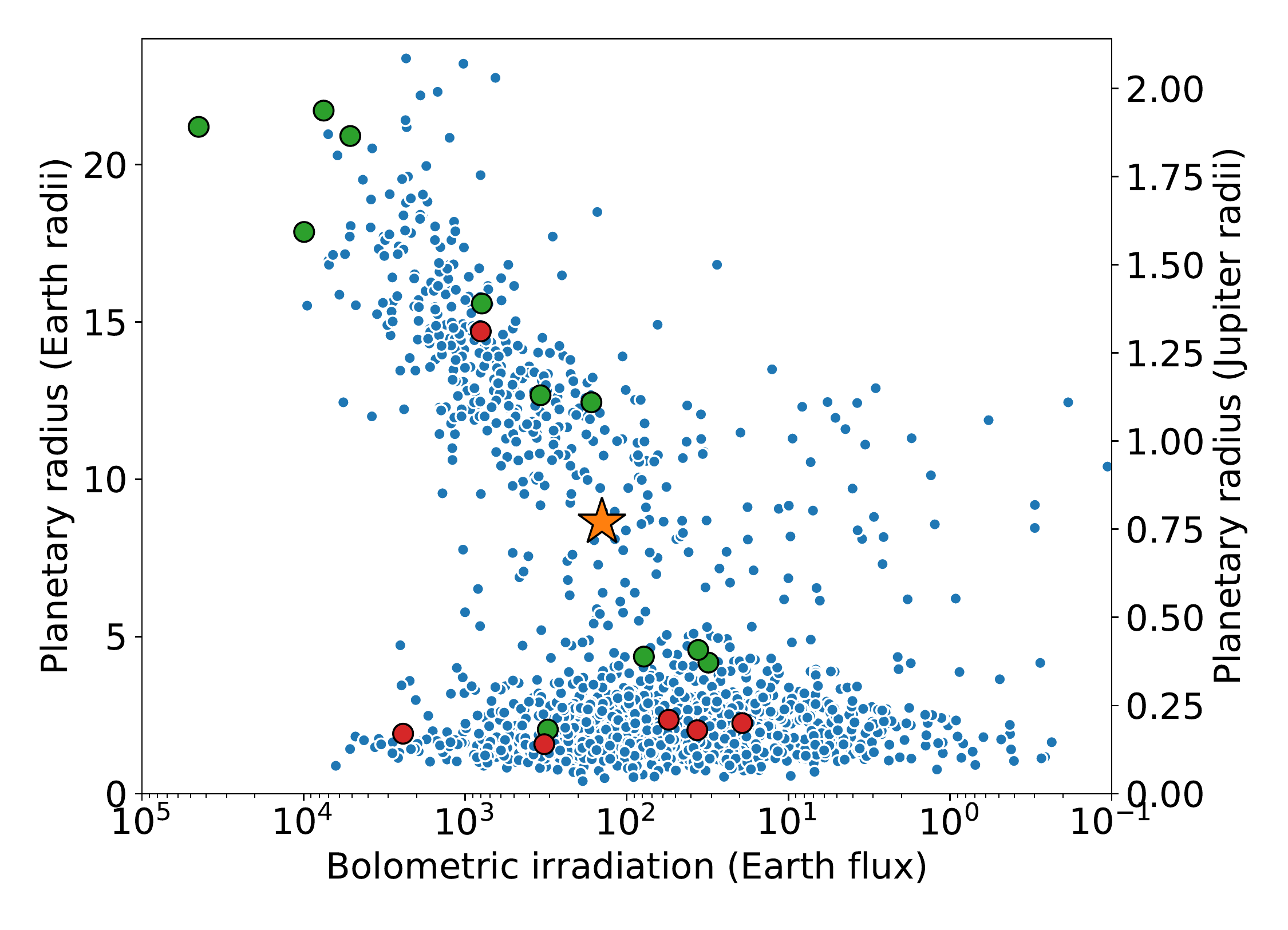}
\caption{Population of exoplanets with known radius and bolometric irradiation (taken from the NASA Exoplanets Archive on 01 Feb 2021). Planets with confirmed atmospheric escape signatures of H, He or metals are shown as green symbols. Non-detections are shown as red symbols. WASP-29~b is represented as a star symbol.}
\label{exopop}
\end{figure}

These signals point to the presence of H-rich, extended atmospheres and mass loss in both ends of the gas giant demographic, but the middle range that includes hot and warm Saturn-sized planets remains poorly explored (Fig. \ref{exopop}). Direct and indirect observational evidence has shown that mass loss is one of the most important drivers for atmospheric evolution in short-period Neptune-sized exoplanets, as well as planetary migration and formation \citep[e.g.,][]{2011ApJ...727L..44S, 2013ApJ...763...12B, 2016A&A...589A..75M, 2016MNRAS.455L..96H, 2016ApJ...820L...8M, 2018MNRAS.479.5012O, 2018A&A...620A.147B, 2019ApJ...876...22M, 2020arXiv200310314A}; although, readers can refer to \citet{2017MNRAS.466.1868C}, \citet{2019ApJ...887L..33K}, and \citet{2020arXiv200310272V} for possible counterpoints. Such a conclusion is backed up by a strong theoretical framework as well \citep[e.g.,][]{2007A&A...461.1185L, 2011ApJ...738...59R, 2011A&A...532A...6S,2014ApJ...783...54K, 2015AREPS..43..459T, 2016A&A...586A..75S}. For a detailed review on the subject, readers can look to \citet{2019AREPS..47...67O}. This atmospheric evolution results in close-in Neptunes having short lifetimes, imprinting a feature in the population of known transiting exoplanets known as the hot-Neptune desert \citep{2011ApJ...727L..44S, 2016A&A...589A..75M, 2018A&A...620A.147B}. 

Hot Jupiters have been shown to be too massive to lose a significant fraction of their atmospheres due to photoevaporation \citep[e.g.,][]{2007P&SS...55.1426G, 2007Icar..187..358H, 2011A&A...529A.136E, 2019MNRAS.490.3760A}. However, \citet{2019ApJ...884L..43G} have proposed that ultra-hot Jupiters orbiting early-type stars, such as KELT-9~b, may have lifetimes of less than 1~Gyr due to near-ultraviolet energy deposited through Balmer absorption. The lack of direct observational constraints on mass loss for Saturn-sized planets at the borders of the hot-Neptune desert contributes to the aforementioned poorly explored demographic. By directly detecting and assessing the current atmospheric rate and other properties of a sample of planetary systems, it is possible to tease apart the respective roles of planetary formation, migration, stellar wind, and photoevaporation in shaping the observed demographics of gas giants \citep{2014ApJ...795...65J, 2018MNRAS.479.5012O, 2020arXiv200310272V}.

One of the objectives of the \emph{Hubble Space Telescope} Panchromatic Comparative Exoplanetology (\hst\ PanCET) program is to address questions about atmospheres of gas giant transiting planets at a demographic level. One of the targets of this program is WASP-29~b, a low-density and moderately irradiated Saturn-sized planet orbiting every 3.92~d around a K4V-type star first discovered by \citet[][the stellar and planetary parameters are listed in Table \ref{w29_param}]{2010ApJ...723L..60H}. A previous attempt at studying the atmosphere of WASP-29~b using ground-based, low-resolution ($R < 1000$) transmission spectroscopy between 515 and 720 nm yielded a featureless spectrum \citep{2013MNRAS.428.3680G}. In particular, the lack of a broad Na feature indicates that either the planet does not have neutral Na in its upper atmosphere, or that clouds and hazes obscure the broad wings of the feature, resulting in a narrow absorption that is not detectable at low spectral resolution.

\begin{table}
\caption{Stellar and planetary parameters of the WASP-29 system.}
\label{w29_param}
\centering
\begin{tabular}{l c c}
\hline\hline
\multicolumn{2}{c}{Stellar parameters of WASP-29} & Ref. \\
\hline
Radius & $0.813^{+0.033}_{-0.015}$ R$_\odot$ & (a) \\
Mass & $0.825 \pm 0.033$ M$_\odot$ & (b) \\
Eff. temperature & $4731^{+44}_{-95}$ K & (a) \\
Proj. rot. velocity & $1.5 \pm 0.6$ km s$^{-1}$ & (b) \\
Radial velocity & $24.48 \pm 0.43$ km s$^{-1}$ & (a) \\
Distance & $87.82 \pm 0.31$ pc & (a) \\
Sp. type & K4V & (c) \\
\hline
\multicolumn{2}{c}{Planetary parameters of WASP-29~b} & Ref. \\
\hline
Radius & $0.792^{+0.056}_{-0.035}$ R$_\mathrm{Jup}$ & (b) \\
Mass & $0.245^{+0.023}_{-0.022}$ M$_\mathrm{Jup}$ & (b) \\
Density & $0.61 \pm 0.12$ g cm$^{-3}$ & (b) \\
Eq. temperature & $970.0^{+32.0}_{-31.0}$ K & (c) \\
Orbital period & $3.9227186 \pm 0.0000068$ d & (b) \\
Semi-major axis & $0.04566^{+0.00060}_{-0.00062}$ au & (b) \\
Ref. time (BJD) & $2455830.18886 \pm 0.00016$ & (b) \\
Transit duration & $2.6486^{+0.0170}_{-0.0151}$~h & (b) \\
Orbital inclination & $88.80 \pm 0.70 \deg$ & (b) \\
Eccentricity & < 0.059 & (b) \\
\hline
\end{tabular}
\tablebib{(a) \citet{2018A&A...616A...1G}, (b) \citet{2017A&A...602A.107B}, (c) \citet{2013MNRAS.428.3680G}.}
\end{table}

In this paper, we report on the data analysis and results of several transits of WASP-29~b observed with \hst\ in FUV, aiming to detect and study its upper atmosphere. In particular, we are interested in addressing the following open issues: i) how the high-energy environment around WASP-29~b influences the observable signatures of atmospheric escape; ii) what escape rates should we expect for hot Saturns similar to WASP-29~b; and iii) whether it shows signals of hydrodynamic escape such as hot Jupiters and warm Neptunes. This manuscript has the following structure: in Sect. \ref{methods} we describe the \hst\ observations and data reduction; in Sections \ref{results} and \ref{cos_analysis} we present the analysis of the flux time series and the search for planetary signals in transmission; in Sect. \ref{he_results} we discuss the reconstruction of the unobservable high-energy spectrum of WASP-29; and in Sect. \ref{conclusions} we present the conclusions of this work.

\section{Description of observations}\label{methods}

We observed six transits of WASP-29~b in the \hst\ PanCET program (GO-14767, PIs: D. Sing and M. L\'opez-Morales), of which three were performed with the Cosmic Origins Spectrograph (COS; Visits A, B, and C) and three with the Space Telescope Imaging Spectrograph (STIS; Visits D, E, and F). Each transit was observed in a single visit with five orbits in time-tag mode. The uncertainty of the transit midpoint time propagated to our latest exposure (21 August 2019) is 7.6 min, based on the ephemeris of \citet{2017A&A...602A.107B}. The COS observations were set to spectroscopic element G130M centered at 1291 \AA\ and a circular aperture with diameter 2.5 arcsec, yielding wavelength ranges [1134, 1274] \AA\ and [1290, 1429] \AA. The STIS observations were set to spectroscopic element G140M centered at 1222 \AA\ and a slit with dimensions $52 \times 0.1$ arcsec (the observations log is shown in Table \ref{log}), yielding a wavelength range [1194, 1248] \AA. 

\begin{table}
\caption{Observations log of WASP-29 in the \hst\ PanCET program.}
\label{log}
\centering
\begin{tabular}{l c c c c}
\hline\hline
\multirow{2}{*}{Visit} & \multirow{2}{*}{Orbit} & Start time & Exp. time & Phase \\
& & (UT) & (s) & (h) \\
\hline
\multicolumn{5}{c}{COS spectroscopy} \\
\hline
\multirow{5}{*}{A} & 1 & 2018-05-19 15:49:32 & 2182.144 & $-3.54$ \\
& 2 & 2018-05-19 17:08:21 & 2706.112 & $-2.01$ \\
& 3 & 2018-05-19 18:43:40 & 2706.112 & $-0.42$ \\
& 4 & 2018-05-19 20:20:47 & 2706.176 & $+1.20$ \\
& 5 & 2018-05-19 22:01:53 & 2706.176 & $+2.85$\\
\hline
\multirow{5}{*}{B} & 1 & 2018-07-17 12:30:51 & 2182.112 & $-3.97$ \\
& 2 & 2018-07-17 13:56:29 & 2706.144 & $-2.47$ \\
& 3 & 2018-07-17 15:31:55 & 2706.176 & $-0.88$ \\
& 4 & 2018-07-17 17:09:16 & 2706.112 & $+0.74$\\
& 5 & 2018-07-17 18:48:57 & 2706.144 & $+2.40$ \\
\hline
\multirow{5}{*}{C}& 1 & 2018-11-04 09:01:12 & 2182.208 & $-4.48$ \\
& 2 & 2018-11-04 10:29:04 & 2706.208 & $-3.09$ \\
& 3 & 2018-11-04 12:04:32 & 2706.208 & $-1.51$ \\
& 4 & 2018-11-04 13:41:15 & 2706.176 & $+0.11$ \\
& 5 & 2018-11-04 15:20:13 & 2706.208 & $+1.80$ \\
\hline
\multicolumn{5}{c}{STIS spectroscopy} \\
\hline
\multirow{5}{*}{D} & 1 & 2019-08-09 20:50:56 & 1751.159 & $-3.90$ \\
& 2 & 2019-08-09 22:14:23 & 1983.173 & $-2.48$ \\
& 3 & 2019-08-09 23:49:48 & 1983.189 & $-0.89$ \\
& 4 & 2019-08-10 01:25:12 & 1983.179 & $+0.70$ \\
& 5 & 2019-08-10 03:00:36 & 1983.195 & $+2.29$ \\
\hline
\multirow{5}{*}{E} & 1 & 2019-08-17 17:54:04 & 1751.138 & $-3.14$ \\
& 2 & 2019-08-17 19:21:08 & 1983.132 & $-1.66$ \\
& 3 & 2019-08-17 20:56:28 & 1983.115 & $-0.07$ \\
& 4 & 2019-08-17 22:31:47 & 1983.187 & $+1.52$ \\
& 5 & 2019-08-18 00:07:06 & 1983.183 & $+3.11$ \\
\hline
\multirow{5}{*}{F} & 1 & 2019-08-21 15:37:18 & 1751.147 & $-3.57$ \\
& 2 & 2019-08-21 17:04:49 & 1983.181 & $-2.08$ \\
& 3 & 2019-08-21 18:40:08 & 1983.015 & $-0.49$ \\
& 4 & 2019-08-21 20:15:27 & 1983.193 & $+1.10$ \\
& 5 & 2019-08-21 21:50:46 & 1983.197 & $+2.69$ \\
\hline
\end{tabular}
\tablefoot{Phases are in relation to the orbit of WASP-29~b.}
\end{table}

\subsection{Data reduction}

\begin{figure}
\includegraphics[width=\hsize]{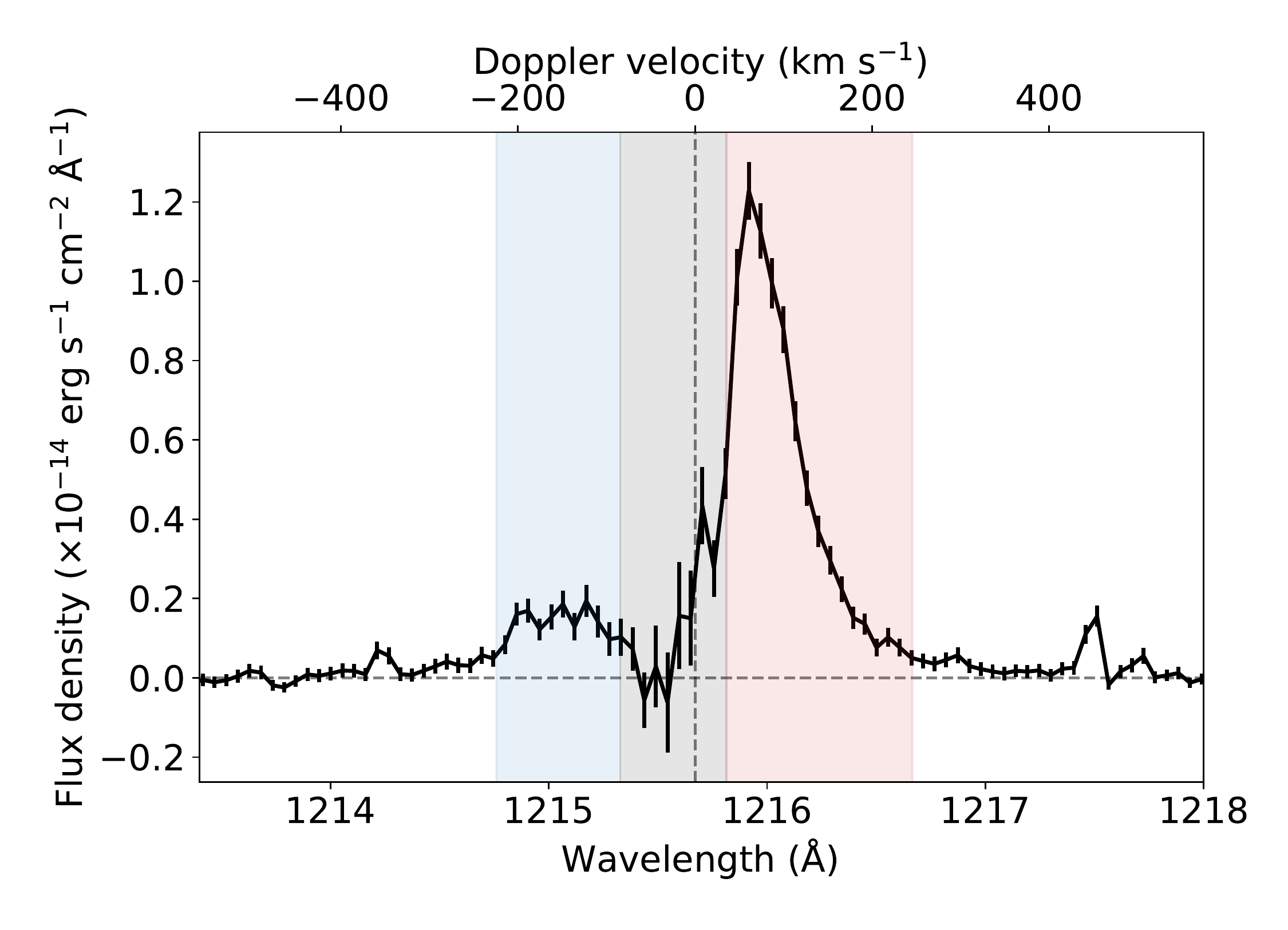}
\caption{Observed \lya\ spectrum of WASP-29 with \hst-STIS in a total exposure time of 8 hours. The gray band represents the wavelength span with strong geocoronal contamination (removed in data reduction). The blue and red regions represent the respective blue and red wings. The Doppler velocities are in the stellar rest frame.}
\label{w29_lya}
\end{figure}

The COS data were reduced automatically by the instrument pipeline. The STIS data required some custom reduction due to strong background contamination in some exposures. The custom STIS reduction consists in using the pipeline flat-fielded frames to extract the spectra and perform background removal. We extracted the spectra following a similar procedure as in \citet{2020A&A...634L...4D}, in which we manually set the cross-dispersion positions of the spectra and the background; the extraction box width was kept the same as in the pipeline reduction. We determined, by visual inspection, the spectrum cross-dispersion positions to be $y = 189$ px and set the background to be centered 20 px away from the spectrum. We combined all STIS exposures to produce a high signal-to-noise \lya\ spectrum, which is shown in Fig. \ref{w29_lya}. We used the time-tag information of the STIS and COS exposures to produce four subexposures per orbit. 

\subsection{Treatment of uncertainties}

WASP-29 is a faint star in FUV, which means the observed count rates are limited to just a few photons per exposure (except for the \lya\ line, which is the brightest feature in this wavelength region). In this regime, the uncertainties of the flux measurements do not follow a Gaussian distribution, but rather a Poisson distribution. Both pipelines of STIS and COS (respectively {\tt calstis} and {\tt calcos}) can calculate the uncertainties of FUV observations using Poisson confidence intervals. In particular, {\tt calcos} produces confidence intervals calculated with the methodology outlined in \citet{1986ApJ...303..336G}, while {\tt calstis} returns confidence intervals assuming Gaussian noise\footnote{\footnotesize{Except for time-tag split data processed with the {\tt inttag} method of {\tt calstis}, which also produces uncertainties based on the methodology of \citet{1986ApJ...303..336G}.}}. 

Due to the non-Gaussian nature of these uncertainties, care has to be taken when propagating them. This is particularly important when calculating the final uncertainties of, for example, transmission spectra, fluxes inside a pass band that is several pixels wide, or combined exposures to produce a higher signal-to-noise spectrum. It is unclear if this is the same effect seen by \citet{2017A&A...599A..75W}, who reported overestimated flux uncertainties produced by the {\tt calcos} pipeline.

In this work, when computing light curves of spectral lines and combining spectra from COS, we work with the raw or net counts and, at the final step, convert them to the physical flux in order to perform a proper treatment of uncertainties. The 1$\sigma$ uncertainties are calculated from the confidence intervals of a Poisson distribution based on the raw number of counts registered in the detector, which includes background sources. The conversion between counts and physical flux is performed using the tabulated sensitivity of the detector. The STIS \lya\ spectra possess count rates at least one order of magnitude higher than those of COS, so the assumption of Gaussian uncertainties is appropriate for STIS spectra.

\subsection{Search for instrumental systematics}\label{jitter}

The engineering jitter files of \hst\ contain telemetry and pointing information that can be used to understand potential systematic trends in long-slit observations with STIS. This method was first used by \citet{2019AJ....158...91S} and we apply our implementation of the technique. The main source of systematic trends in STIS data is slit losses, which are caused by two effects: i) changes in the position of the target star in the slit, both in the dispersion and cross-dispersion directions, and ii) changes in the point-spread function (PSF) seen by the detector due to thermal variations of the telescope -- also known as breathing. Previous ultraviolet studies with \emph{HST} have shown that these systematic trends are more important the brighter is the observed star \citep[e.g.,][]{2017A&A...597A..26B, 2018A&A...615A.117B, 2019AJ....158...91S}, and are not present in COS spectroscopy \citep{2019A&A...629A..47D}.

The pointing of the telescope can be traced by the right ascension and declination data present in the jitter data, but what matters in our case is the position of the target in relation to the slit. While we do not know exactly where our target is in relation to the slit at a sub-pixel level, we can gather this information from the jitter data of the guide stars. Usually, \hst\ observations are performed with two guide stars, called dominant and roll; their positions in relation to the V$_2$ and V$_3$ axes of the telescope are traced to sub-pixel levels; since the slit is not aligned to the reference axes of the telescope, we need to rotate the dominant and roll position vectors to the slit reference frame -- we call these new vectors V$_{\rm d}$ (dispersion direction) and V$_{\rm s}$ (spatial, or cross-dispersion direction). 

\begin{table*}
\caption{Correlation between \lya\ red wing fluxes and jitter data.}
\label{jitter_corr}
\centering
\begin{tabular}{l | c c | c c | c c}
\hline\hline
    & \multicolumn{2}{c |}{Visit D} & \multicolumn{2}{c |}{Visit E} & \multicolumn{2}{c}{Visit F} \\
    & Spearman-$r$ & $p$-value & Spearman-$r$ & $p$-value & Spearman-$r$ & $p$-value \\
\hline
V$_{\rm d, dom}$  & $-0.547$ & $0.012$ & $+0.223$ & $0.346$ & $-0.293$ & $0.210$ \\
V$_{\rm d, roll}$ & $-0.505$ & $0.023$ & $+0.206$ & $0.384$ & $-0.200$ & $0.398$ \\
V$_{\rm s, dom}$  & $-0.030$ & $0.900$ & $-0.260$ & $0.269$ & $-0.186$ & $0.431$ \\
V$_{\rm s, roll}$ & $+0.164$ & $0.490$ & $+0.268$ & $0.254$ & $+0.006$ & $0.980$ \\
Longitude         & $-0.090$ & $0.705$ & $+0.147$ & $0.535$ & $+0.268$ & $0.254$ \\
Latitude          & $-0.017$ & $0.945$ & $-0.376$ & $0.102$ & $-0.134$ & $0.574$ \\
Right ascension   & $+0.365$ & $0.113$ & $-0.129$ & $0.587$ & $+0.487$ & $0.029$ \\
Declination       & $-0.564$ & $0.010$ & $+0.202$ & $0.394$ & $-0.215$ & $0.363$ \\
\hline
\end{tabular}
\end{table*}

In order to infer if the jitter trends cause slit losses in the STIS exposures, we searched for correlations between the observed \lya\ red wing emission fluxes of the time-tag split data and the jitter data using the Spearman-$r$ coefficient \citep{10.2307/1412159}. The results are shown in Table \ref{jitter_corr}. Despite Visit E displaying the strongest jitter trends, the correlation with the \lya\ fluxes are weak, with $p$-values above 0.10. In relative terms, the strongest (anti-)correlations are seen in Visit D, where the guide star V$_{\rm d}$ vector and the declination have Spearman-$r$ coefficients around 0.50 in modulus. Nevertheless, in absolute terms, these coefficients are below the threshold of 0.75 in modulus to be considered significant. Thus we conclude that there are negligible jitter trends in our STIS dataset. 

We also searched for breathing effects by phase folding the subexposures to the orbit of \hst\ and comparing the data to Fourier decomposition models \citep[as in][]{2017A&A...597A..26B}. We evaluated the likelihood of breathing effects by comparing the Bayesian Information Criterion (BIC) of models with varying orders fitted to the observed data. In all visits a flat model with no systematics is favored against higher order models: $\Delta {\rm BIC} = 20$ for Visit D, $\Delta {\rm BIC} = 9$ for Visit E, and $\Delta {\rm BIC} = 12$ for Visit F.

\section{\lya\ emission time series}\label{results}

The \lya\ emission line (\ion{H}{I}) is the strongest feature in the far-ultraviolet spectrum of cool stars \citep{2017ARA&A..55..159L} and plays an important role in the photochemistry of planetary atmospheres \citep{2005AsBio...5..706S}. For stars other than the Sun, the \lya\ emission is partially or completely absorbed by the interstellar medium (ISM). For the nearest stars, the core of the line is inaccessible, but its wings can still be observed, yielding the usual double-peaked profile (Fig. \ref{w29_lya}). The shape of this observed profile will depend on the properties of the intrinsic stellar emission, the physical properties of the ISM, and the radial velocity of the observed star.

\begin{figure*}
\centering
\includegraphics[width=\hsize]{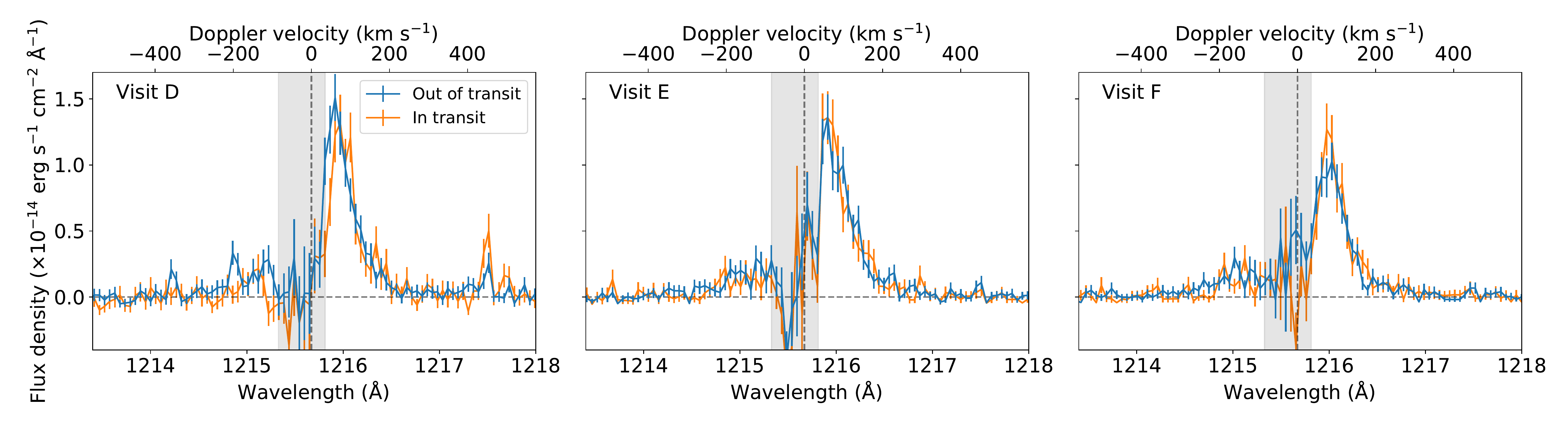}
\caption{In- and out-of-transit \lya\ spectra of WASP-29 in Visits D, E, and F. The shaded region represents the interval where there is strong geocoronal contamination. We do not detect an obvious in-transit absorption signal.}
\label{lya_transit_profiles}
\end{figure*}

Atmospheric signatures in hot transiting exoplanets have been previously detected as a large (> 10\%) absorption in the \lya\ line, which traces exospheric \ion{H}{I} atoms \citep[e.g.,][]{2003Natur.422..143V, 2012A&A...543L...4L, 2015Natur.522..459E, 2018A&A...620A.147B}. Such a signal indicates the presence of a H-rich exospheric cloud being accelerated by radiation pressure and interactions with the stellar wind through charge exchange. We searched for a similar feature in the \lya\ transit time series of WASP-29. At first, this search was performed by visual inspection of the \lya\ profiles (Fig. \ref{lya_transit_profiles}). Upon concluding that there was no obvious signal, we performed a light curve analysis of the fluxes in the blue and red wing bandpasses. We define these bandpasses by the limits of the geocoronal contamination ([-60, +60]~km~s$^{-1}$ in the heliocentric restframe, or [-84, +35]~km~s$^{-1}$ in the stellar restframe) and the approximate limits where the stellar flux is above the 1$\sigma$ noise floor of the instrument. This yields the blue and red wing bandpass limits as [-224, -84]~km~s$^{-1}$ and [+35, +245]~km~s$^{-1}$, respectively.

In a single STIS exposure, we can measure the \lya\ blue wing flux with a mean precision of 41.6\%, 30.4\%, and 29.1\%, respectively for visits D, E, and F. In the case of the red wing fluxes, the mean precision values are 10.4\%, 9.5\%, and 9.9\%, respectively. Due to potential stellar flux modulation in the time scales that separate the visits, we normalize each time series by the out-of-transit flux. This baseline is measured as the average flux in orbits 1, 2, and 5 of each visit. We fit the normalized light curves to transit models calculated with \texttt{batman} \citep{2015PASP..127.1161K} and estimate the uncertainties of the fit using the Markov chain Monte Carlo ensemble sampler implementation of \texttt{emcee} \citep{2013PASP..125..306F}. We fix the orbital parameters of WASP-29~b to the values shown in Table \ref{w29_param} and vary only the depth $(R_{\rm p} / R_{\rm *})^2$.

\begin{figure*}
\centering
\begin{tabular}{cc}
\includegraphics[width=0.44\hsize]{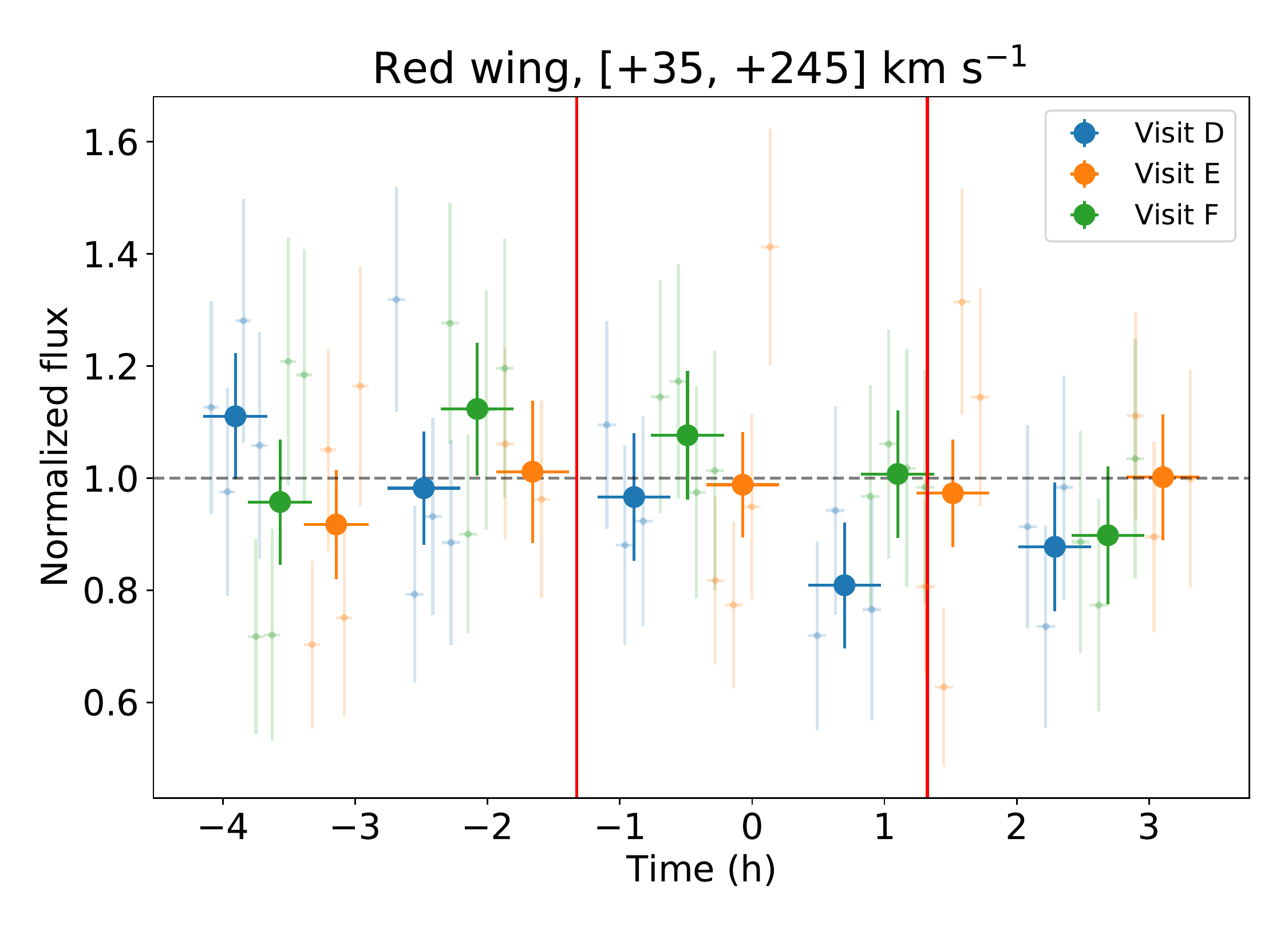} & \includegraphics[width=0.44\hsize]{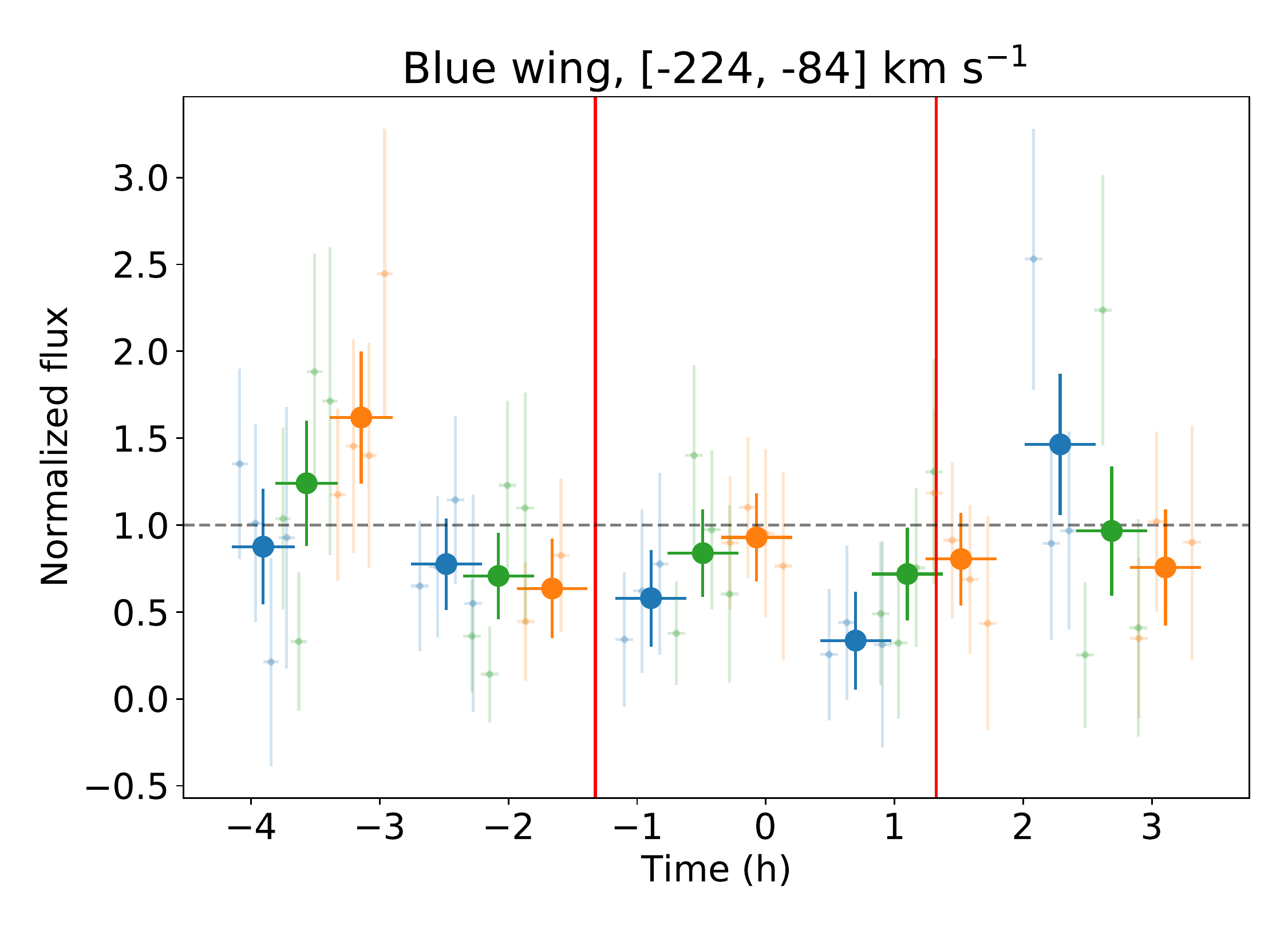} \\
\end{tabular}
\caption{Normalized \lya\ light curves of WASP-29. The vertical red lines represent the times of first and fourth contacts of the transit of WASP-29~b. Each \emph{HST} orbit is represented as a full symbol, while semi-transparent symbols represent sub-exposures within each orbit.}
\label{lya_lc}
\end{figure*}

There is a flux decrease of $\sim$20\% in the \lya\ red wing in Visit D, a feature that is seen in the in- and out-of-transit \lya\ profiles in Fig. \ref{lya_transit_profiles}, and the light curve of the red wing in Fig. \ref{lya_lc}. For comparison, GJ~436 shows an episodic red wing in-transit absorption as well \citep{2019A&A...629A..47D}, but its origin remains unknown. Since this feature is not repeated in the other STIS observations of our program, we deem it unlikely to arise from the exosphere of WASP-29~b. Since the absorption takes place near the core of the \lya\ emission of WASP-29, it is possible that the decrease in flux can be caused by stellar activity modulation. Similar to the case of HD~97658 in \citet{2017A&A...597A..26B}, stellar activity is expected to modulate the wavelength region nearest to the line core, which is formed at higher temperatures in the stellar corona, than its wings.

We did not find evidence to support \lya\ flux absorption during the transit of WASP-29~b. Using the technique described above and by combining all the visits, we rule out any in-transit absorption depth larger than 49\% (92\%) in the blue wing at 1$\sigma$ (3$\sigma$) confidence. For the red wing light curves, we rule out any in-transit absorption signals above 9\% (19\%) at 1$\sigma$ (3$\sigma$) confidence.

For comparison, the \ion{H}{I} detection in HD~209458~b showed an in-transit depth of $15\% \pm 4\%$ in two bandpasses near the line core, between Doppler velocities -130 km s$^{-1}$ and +100 km s$^{-1}$ \citep[excluding the wavelength interval with geocoronal emission contamination;][]{2003Natur.422..143V}. The hot-Jupiter HD~189733~b displays \ion{H}{I} transit depths of $14.4\% \pm 3.6\%$ between Doppler velocities -230 km s$^{-1}$ and -140 km s$^{-1}$ \citep{2013A&A...551A..63B}. The warm Neptunes GJ~436~b and GJ~3470~b display \lya\ transit depths around 50\% in the Doppler velocity interval [-120, -40]~km~s$^{-1}$ \citep{2015Natur.522..459E, 2018A&A...620A.147B}.

\section{Metallic species time series}\label{cos_analysis}

\begin{figure*}
\centering
\includegraphics[width=\hsize]{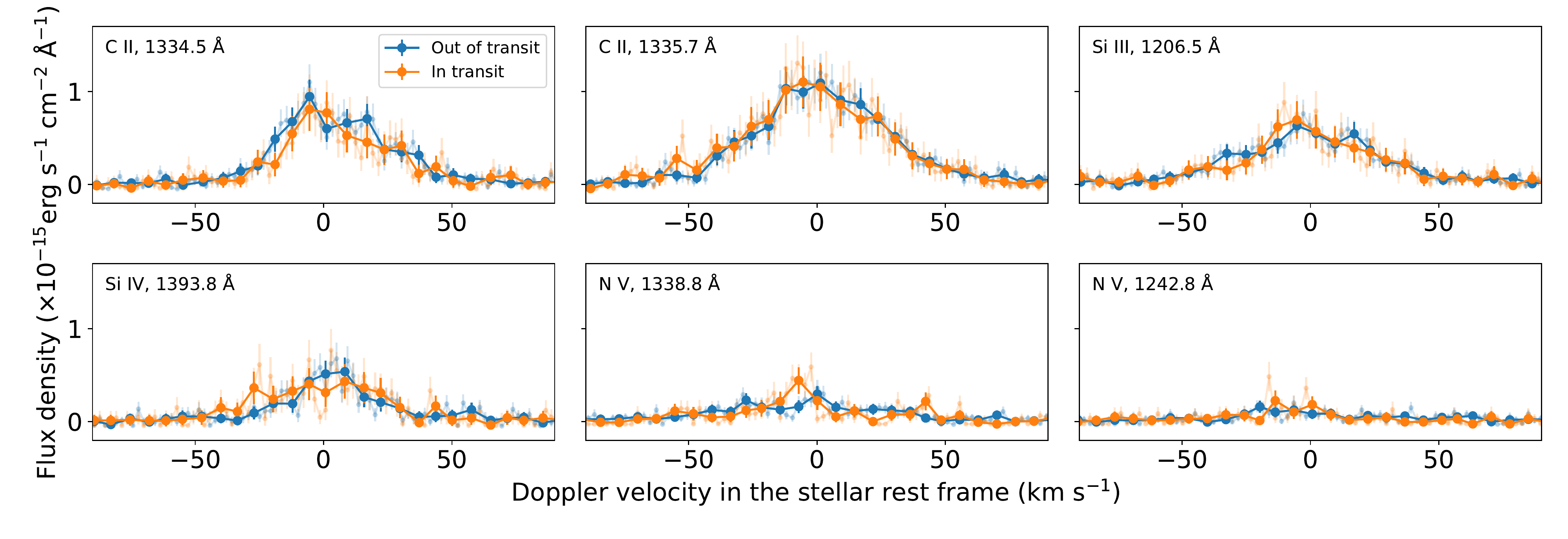}
\caption{In- and out-of-transit metallic spectral lines of WASP-29 in Visits A, B, and C.}
\label{cos_profiles}
\end{figure*}

\begin{figure*}
\centering
\begin{tabular}{cc}
\includegraphics[width=0.44\hsize]{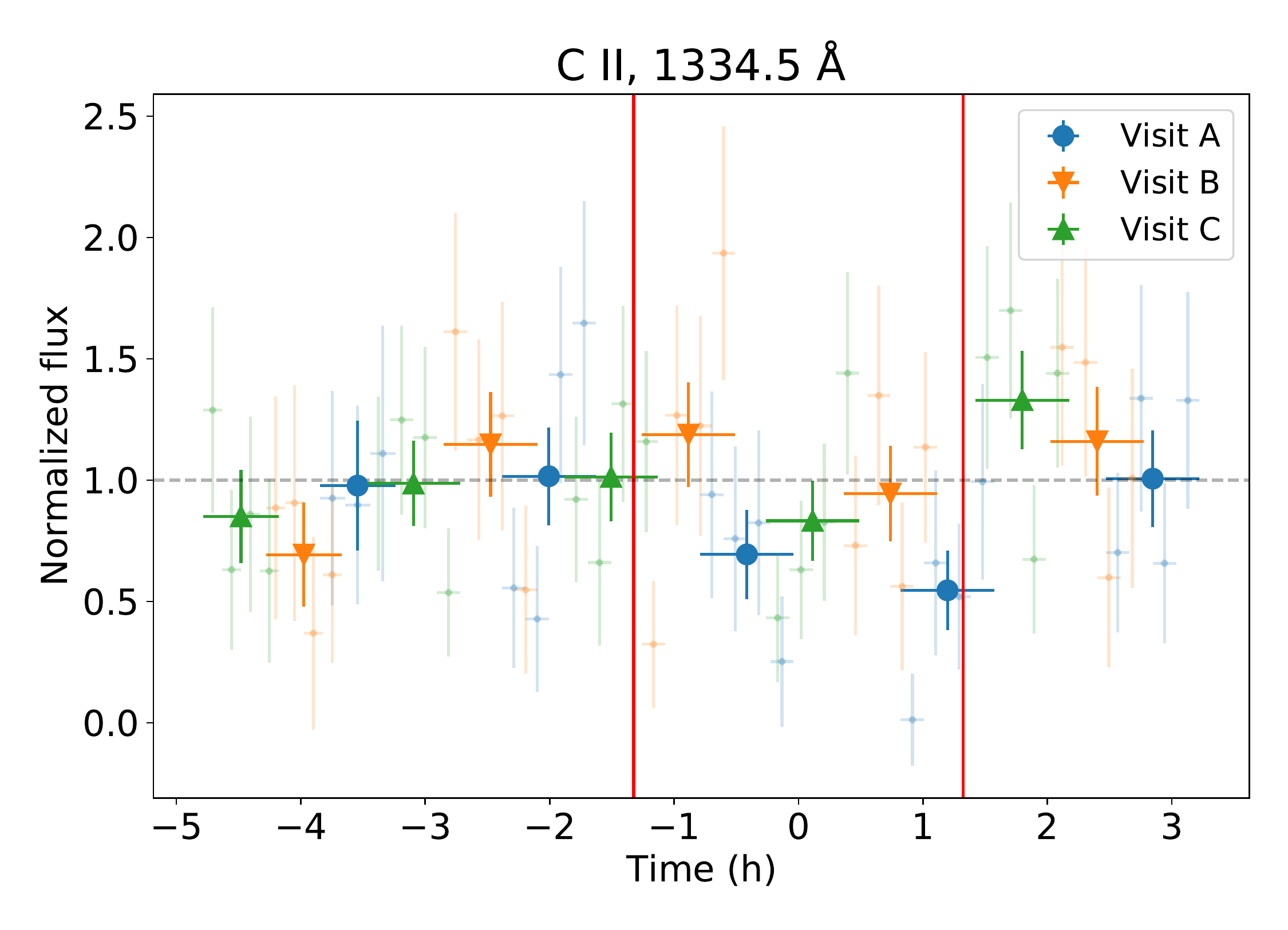} & \includegraphics[width=0.44\hsize]{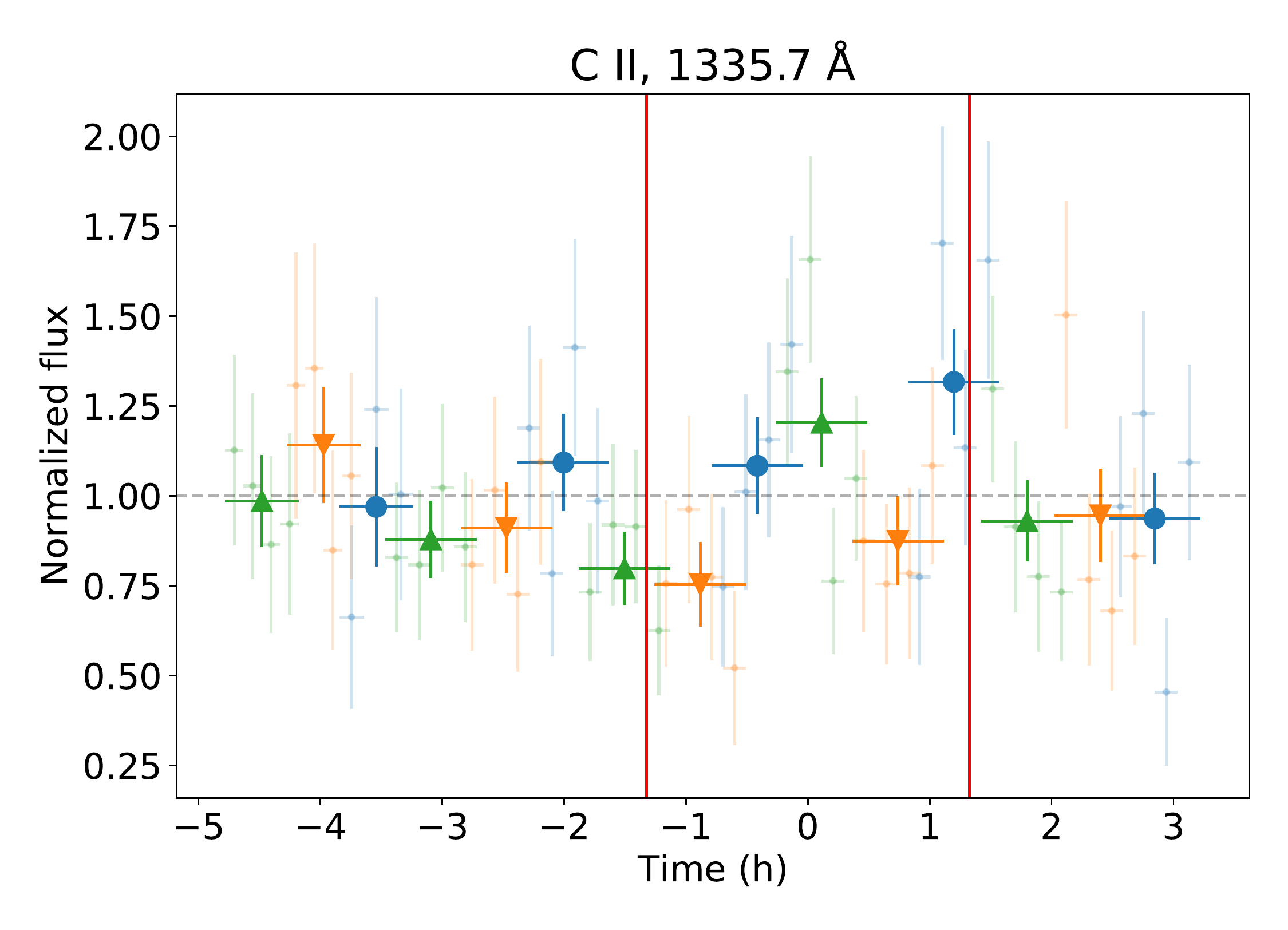} \\
\includegraphics[width=0.44\hsize]{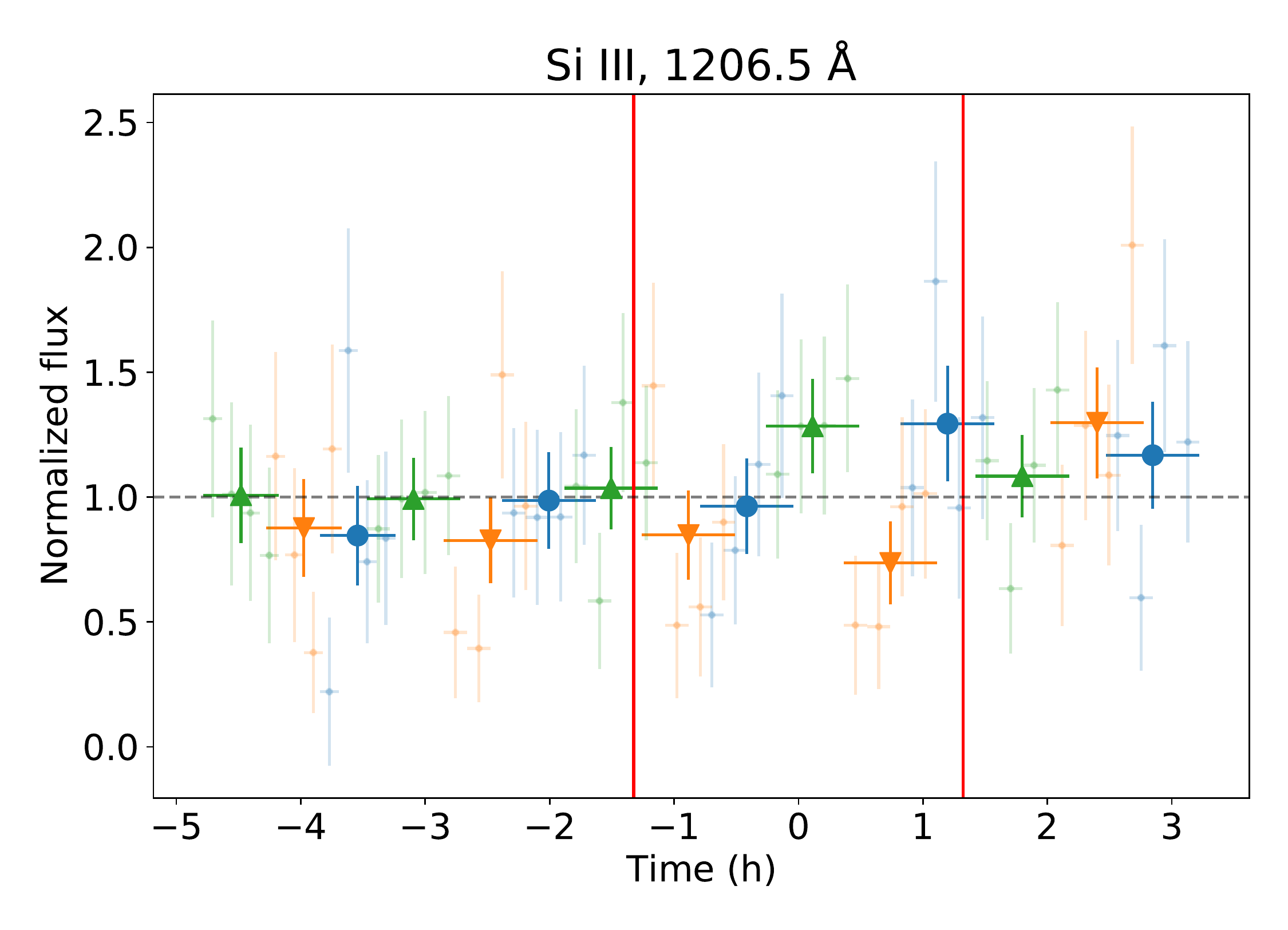} & \includegraphics[width=0.44\hsize]{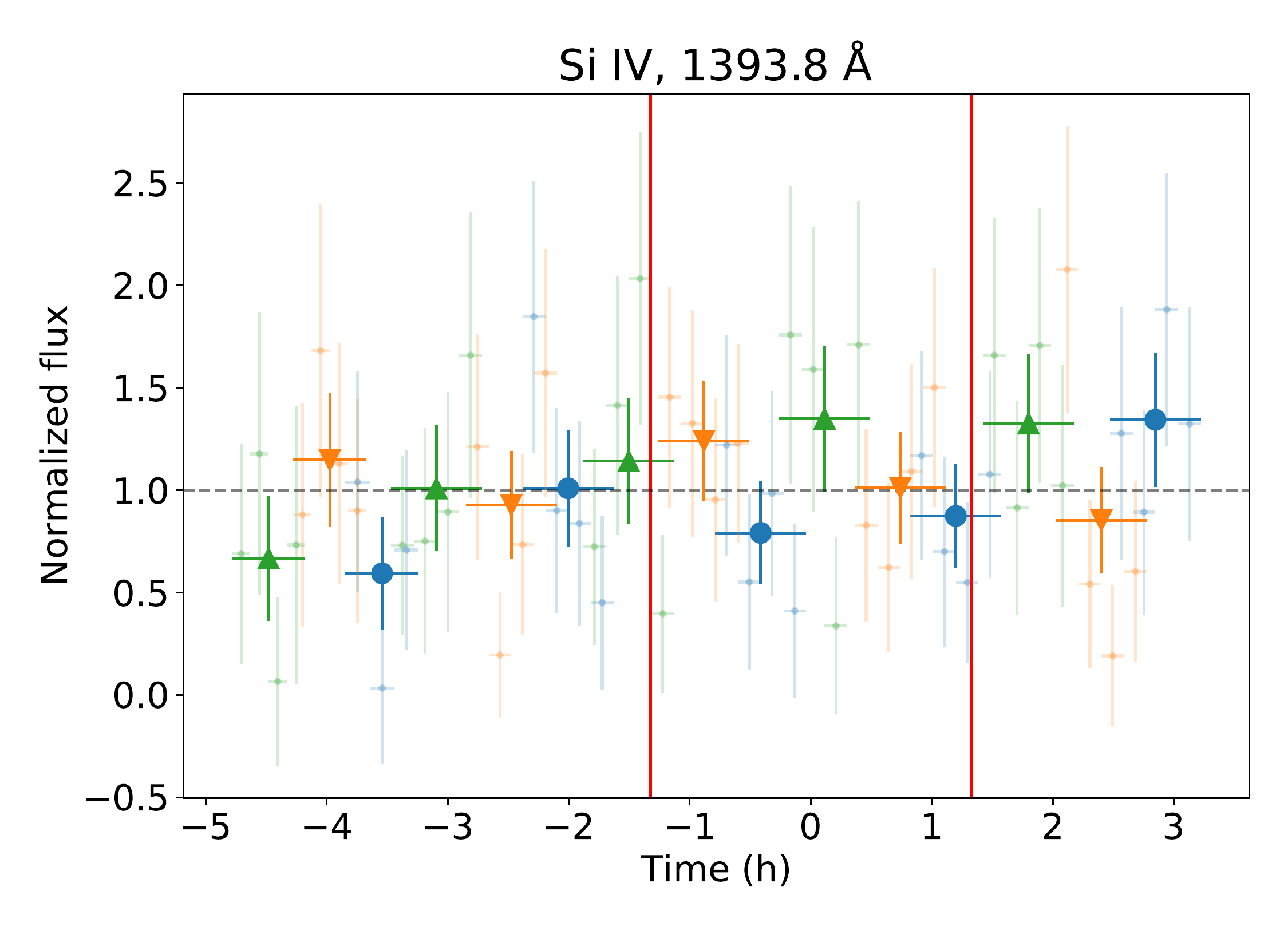} \\ \includegraphics[width=0.44\hsize]{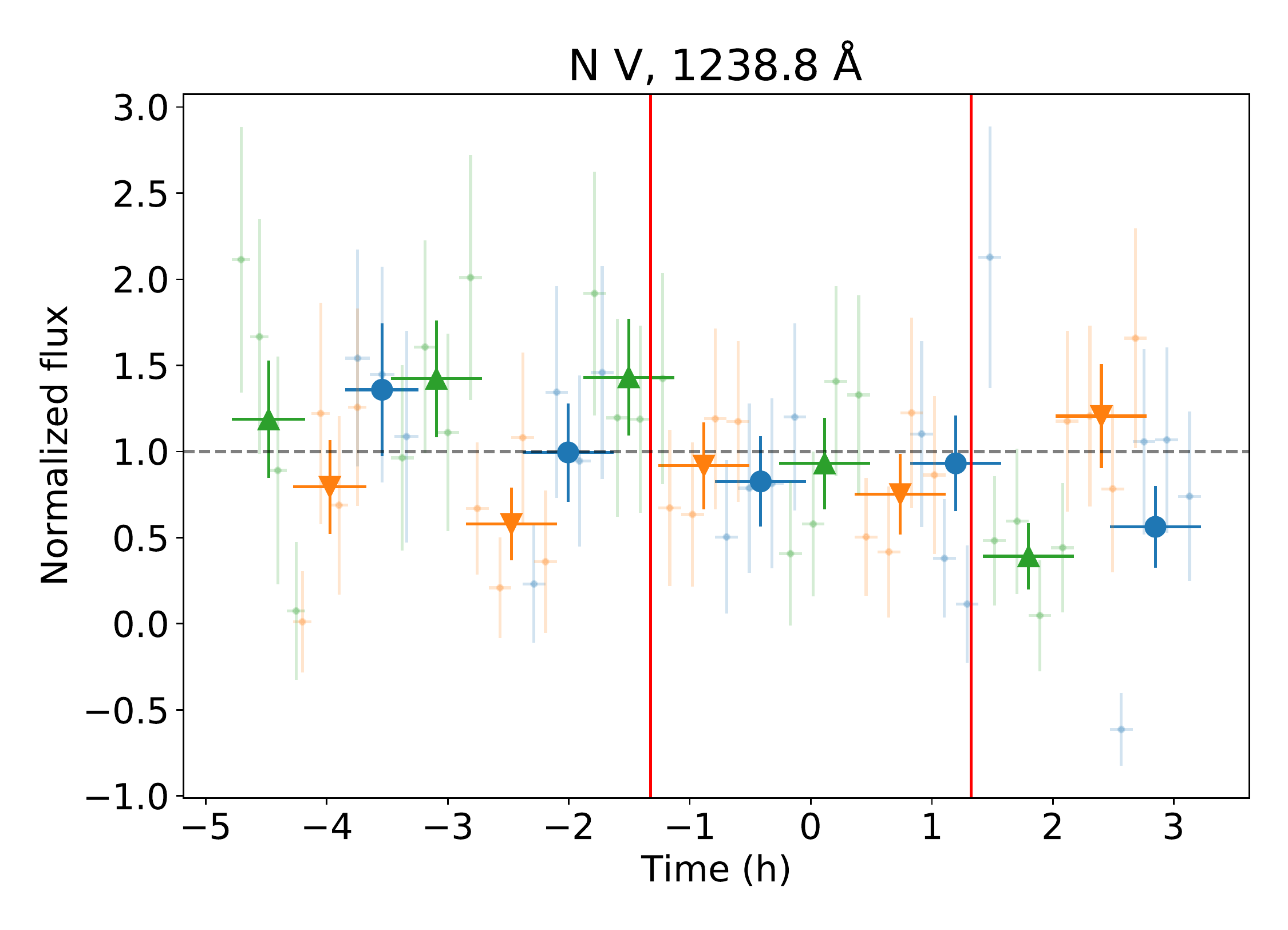} & \includegraphics[width=0.44\hsize]{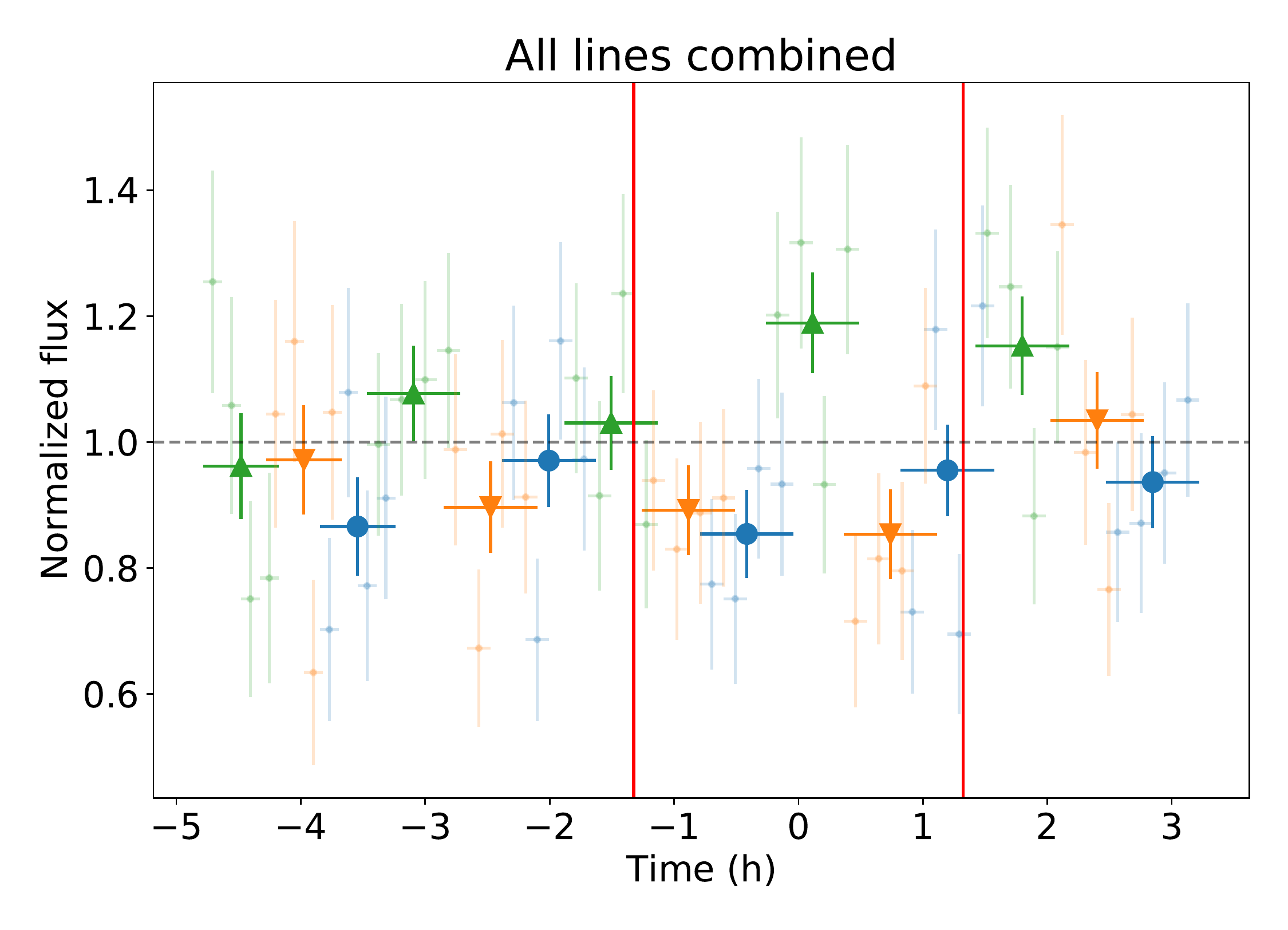}\\
\end{tabular}
\caption{Light curves of the metallic species emission lines of WASP-29 in the COS visits. The vertical red lines represent the times of first and fourth contacts of the transit of WASP-29~b. Each \emph{HST} orbit is represented as a full symbol, while semi-transparent symbols represent sub-exposures within each orbit.}
\label{cos_lc}
\end{figure*}

To date, all detections of exospheric metals in transiting planets have been found in the hot Jupiters HD~209458~b \citep{2004ApJ...604L..69V, 2010ApJ...717.1291L, 2013A&A...560A..54V}, HD~189733~b \citep{2013A&A...553A..52B}, WASP-12~b \citep{2010ApJ...714L.222F} and WASP-121~b \citep{2019AJ....158...91S}. \citeauthor{2019AJ....158...91S} argue that exospheric metals in WASP-121~b can be observed in transmission when the exobase extends beyond the Roche lobe, allowing heavier species in the upper atmosphere to escape the planet. A search for such features in GJ~436~b, which benefited from a more extensive dataset covering several years, resulted in non-detections only \citep{2017A&A...605L...7L, 2017ApJ...834L..17L, 2019A&A...629A..47D}. We perfomed a similar analysis for the three transits of WASP-29~b; the host star is fainter than GJ~436, thus we could analyze fewer spectral lines in the COS spectrum. The COS profiles of \ion{C}{II}, \ion{Si}{III}, \ion{Si}{IV}, and \ion{N}{V} are shown in Fig. \ref{cos_profiles}, and their respective transit light curves are shown in Fig. \ref{cos_lc}.

Similarly to the \lya\ analysis, we fit the observed time series with transit light curve models to assess the absorption depth and its statistical significance. We found that the stellar flux in the line that arises from the ground-level state \ion{C}{II} (1334.5 \AA) decreases by $39\%^{+12\%}_{-11\%}$; but we cannot rule out possible stellar variability effects that could mimic planetary signals since a similar "dip" is also seen before the transit between -5 and -4 hours in relation to the transit center. Furthermore, no clear-cut signal is seen in the combined \ion{C}{II} spectra (Fig. \ref{cos_profiles}). For comparison, \citet{2010ApJ...717.1291L} detected an in-transit absorption of $7.8\% \pm 1.3\%$ in the combined flux of the \ion{C}{II} doublet for HD~209458~b using COS spectra; that result matched well the marginal $2\sigma$ detection of \citet{2004ApJ...604L..69V}, which was performed with the STIS spectrograph. \citet{2021arXiv210200203G} reported a 3.4$\sigma$ detection of excited-level \ion{C}{II} during one transit of the super-Earth $\pi$~Men~c that is consistent with ions filling the Roche lobe of the planet and moving away from the star. The decrease in \ion{C}{II} flux that we observe in WASP-29~b is dominated by Visit A only, while Visits B and C are consistent with flat light curves. Since the signal in Visit A is not repeatable, we deem this decrease in flux as unrelated to the planet WASP-29~b.

The normalized light curve of the line arising from the excited-level \ion{C}{II} atom (1335.7 \AA) is suggestive of a W-shaped time series. Such a transit light curve could be indicative of FUV limb-brightening in the stellar host, a well known phenomenon in the Sun \citep{1955Natur.176..652M}. The effect is clearer when we look at the time-tag subexposures, which are shown as semi-transparent points in Fig. \ref{cos_lc}. This behavior would not arise from the shape of the planetary atmosphere, but from the limb-brightened chromosphere and transition region of the host star. However, more observations would be needed to confirm this feature for WASP-29~b. Since the excited-level \ion{C}{II} line is the strongest feature in the COS spectra after \lya , it dominates the flux of the "all lines combined" light curve (lower right panel in Fig. \ref{cos_lc}) and the W-shape remains present. The COS light curves show that there are no obvious signatures of \ion{Si}{III}, \ion{Si}{IV} or \ion{N}{V} ions populating the exosphere of WASP-29~b. We calculated the 1$\sigma$ and 3$\sigma$ upper limits for the in-transit absorption depth and the observed COS fluxes, and the results are shown in Table \ref{cos_results}. 

\begin{table*}
\caption{Observed fluxes of WASP-29 and upper limits of the in-transit absorption depth for WASP-29~b.}
\label{cos_results}
\centering
\begin{tabular}{lccccc}
\hline\hline
\multirow{2}{*}{Species} & Wavelength & Integration range & Flux & \multicolumn{2}{c}{Transit depth} \\
 & (\AA) & (km~s$^{-1}$) & ($\times 10^{-16}$ erg s$^{-1}$ cm$^{-2}$) & $1\sigma$ & $3\sigma$ \\
\hline
\ion{C}{II} & 1334.5 & [-60, +60] & $1.94 \pm 0.11$ & $39\%^{+12\%}_{-11\%}$ & $39\%^{+40\%}_{-31\%}$ \\
\ion{C}{II} & 1335.7 & [-80, +80] & $2.88 \pm 0.12$ & $< 10\%$ & $< 24$\% \\
\ion{Si}{III} & 1206.5 & [-80, +80] & $1.33 \pm 0.07$ & $11\% \pm 9\%$ & $< 40$\% \\
\ion{Si}{IV} & 1393.8 & [-60, +60] & $0.92 \pm 0.07$ & $< 17$\% & $< 49$\% \\
\ion{N}{V} & 1238.8 & [-80, +80] & $0.63 \pm 0.04$ & $< 31$\% & $< 70$\% \\
\hline
\end{tabular}
\end{table*}

\section{The high-energy environment of WASP-29~b}\label{he_results}

The STIS and COS observations provide useful information about the stellar UV irradiation and its impact on the planet upper atmosphere, as well as constraints on the ISM. They also allow us to interpret the non-detections of atmospheric signals and to predict signals that could be observed in other wavelengths. The ISM absorbs a large portion of the stellar emission at short wavelengths, including the \lya\ emission and the extreme ultraviolet (EUV) spectrum, so they need to be reconstructed.

\subsection{Reconstruction of the intrinsic \lya\ line}

We used the STIS observations of WASP-29 to reconstruct its intrinsic \lya\ spectrum following the standard method used in, for example, \citet{2017AJ....154..121B} and \citet{2018A&A...620A.147B}. In short, we fit the observed spectrum to a model of the intrinsic emission line attenuated by ISM absorption (\ion{H}{I} and deuterium), scaled for distance, and convolved with the instrumental response. The model is oversampled in wavelength and rebinned over the STIS spectral table after convolution; the fit yields an estimate of the intrinsic emission and certain properties of the ISM in the line of sight. 

Since no signatures of planetary absorption or stellar variability were found in any of the visits, the reconstruction was performed using all available spectra averaged into one master spectrum per visit. In all visits we excluded from the fit the velocity range [-60, 60]\,km\,s$^{-1}$ (heliocentric rest frame), which is too contaminated by geocoronal emission. We fit the three master spectra together using the Markov chain Monte Carlo (MCMC) software package \textit{emcee} \citep{2013PASP..125..306F}. K dwarfs can show either double-peaked or single-peaked Lyman-$\alpha$ profiles, depending on the presence of a self-reversal\footnote{\footnotesize{A self-reversal in \lya\ happens in some stars, like the Sun, because there is a layer in of absorbing material in the stellar atmosphere between the region where the line is formed and the outermost stellar layers \citep[see, e.g.,][]{1998ApJS..119..105W}.}} in their core \citep[e.g.,][]{2005ApJS..159..118W, 2016ApJ...824..101Y, 2017A&A...602A.106B}. We found that a single-peaked Voigt profile provides the best fit to the WASP-29 intrinsic Lyman-$\alpha$ line. The theoretical intrinsic line profile is defined by its centroid, its temperature (assuming pure thermal broadening), its Voigt damping parameter, and its total flux $F_{\mathrm{Ly-\alpha}}$(1\,au). Except for the total flux, all properties were set to the same value for the three visits. The line centroid was fixed to the systemic velocity of WASP-29 (24.48\,km\,s$^{-1}$, as in Table \ref{w29_param}). The absorption profile of the ISM along the line of sight is common to the three visits, and defined by its column density of neutral hydrogen log$_{10}$\,$N_{\mathrm{ISM}}$(H\,{\sc i}), its Doppler broadening parameter $b_{\mathrm{ISM}}$(H\,{\sc i}), and its heliocentric radial velocity $\gamma_{\mathrm{ISM}/\sun}$. The D\,{\sc i}/H\,{\sc i} ratio was set to 1.5$\times$10$^{-5}$\ \citep[e.g.,][]{2003ApJ...599..297H, 2006ApJ...647.1106L}. We adopted the Lyman-$\alpha$ oscillator strength $f=0.41641$ from the NIST database\footnote{\footnotesize{Available in \url{https://physics.nist.gov/PhysRefData/ASD/lines_form.html}.}}. The Doppler broadening of the ISM toward WASP-29 is not constrained by its saturated Lyman-$\alpha$ line absorption profile, and was fixed to the value of the Local Interstellar Cloud (LIC).

\begin{figure}
\centering
\includegraphics[width=\hsize]{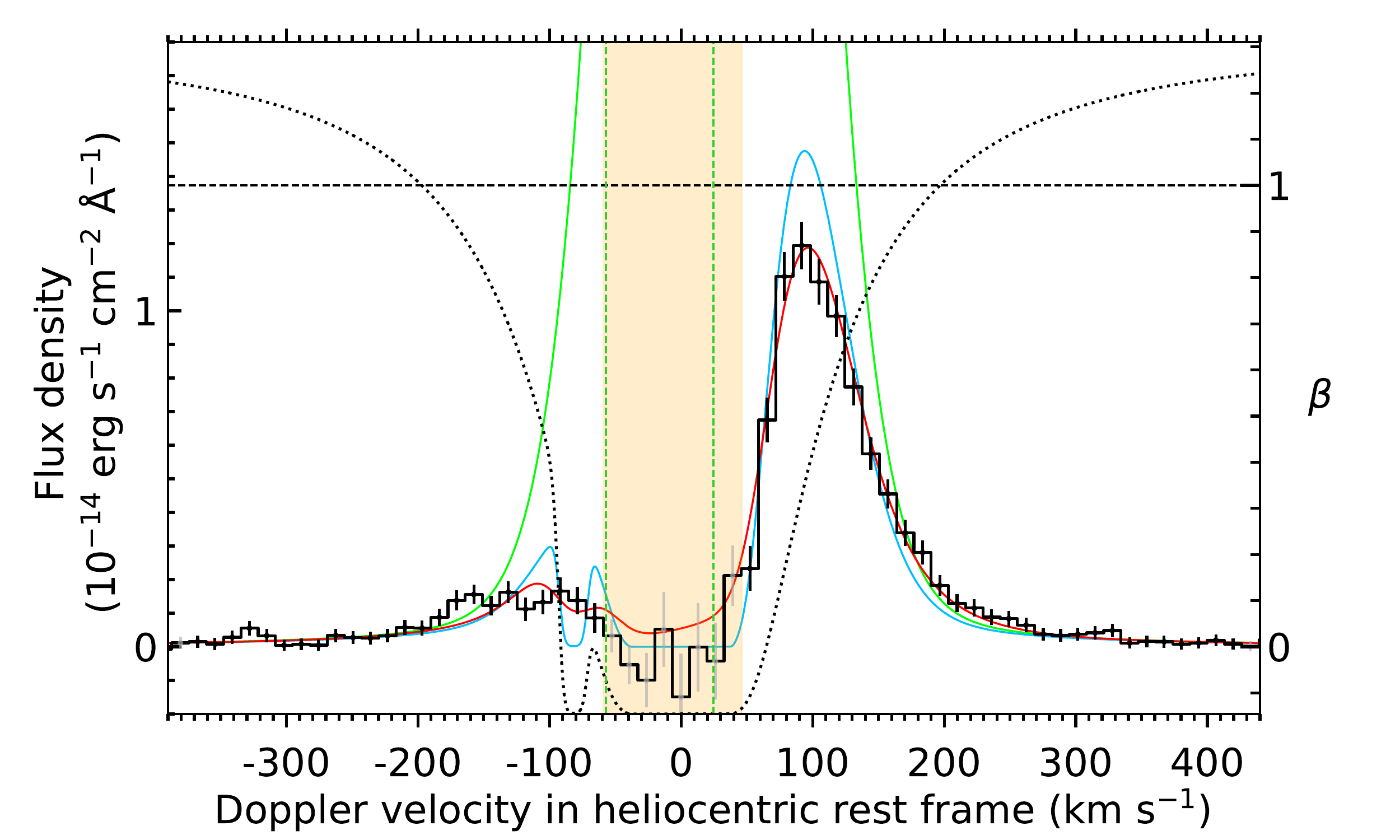}
\caption{Lyman-$\alpha$ line profile of WASP-29 averaged over the three visits. The black histogram shows the observed spectra, fit over points with black error bars. The orange band shows the region strongly contaminated by geocoronal emission to be excluded from the fit. The green full line is the best fit for the intrinsic stellar line profile, and the green dotted line represents the zero velocity in the stellar rest frame. It yields the blue profile after absorption by the interstellar medium, whose profile is plotted as a dotted black line (ISM absorption in the range 0-1 has been scaled to the vertical axis range). The red line shows the line profile fit to the data, after convolution with the STIS instrumental profile. The theoretical intrinsic stellar line profile also corresponds to the profile of the ratio $\beta$ between radiation pressure and stellar gravity, which is shown in the right y-axis.}
\label{lya_rec}
\end{figure}

We obtain a minimum $\chi^2$ of 166 for 151 degrees of freedom (reduced $\chi^2 = 1.1$). The best-fit model for the line averaged over the three visits is shown in Fig.~\ref{lya_rec}. The LISM kinematic calculator\footnote{\mbox{\url{http://sredfield.web.wesleyan.edu/}}}, a dynamical model of the local ISM \citep{2008ApJ...673..283R}, predicts that the line of sight toward WASP-29 crosses the LIC cloud with a heliocentric radial velocity of $-1.29$\,km\,s$^{-1}$. This is in very good agreement with our derived value of $-0.1 \pm 2.6$\,km\,s$^{-1}$. Our best-fit column density (log$_{10}$\,$N_{\mathrm{ISM}}$(H\,{\sc i}) = $18.55 \pm 0.04$\,cm$^{-2}$) is also consistent with the range of values expected for a star at a distance of 87.8\,pc \citep[Fig. 14 in][]{2005ApJS..159..118W}. \\ 

The total flux in the observed and reconstructed Lyman-$\alpha$ lines shows no significant variations between the three visits, strengthening the long-term stability of the stellar UV emission. The average line flux F$_{\mathrm{Ly-\alpha}}$(1\,au) = 17.6$\stackrel{+2.1}{_{-1.6}}$\,erg\,s$^{-1}$\,cm$^{-2}$ yields a rotation period of about 20 days for WASP-29 from the empirical relation in \citet{2016ApJ...824..101Y}. In Fig. \ref{lya_rec}, the $\beta$ parameter is the ratio between radiation pressure and stellar gravity; our reconstruction shows that radiation pressure from WASP-29 Lyman-$\alpha$ line overcomes its gravity by a factor of seven at velocities between [-100, 100]~km~s$^{-1}$ in the stellar rest frame. This is much larger than the factor 3 associated with the K0 dwarfs HD\,189733 and HD\,97658 \citep{2013A&A...551A..63B, 2017A&A...597A..26B}, or even the factor $\sim$4.5 associated with the G0 dwarf HD\,209458 \citep{2013A&A...557A.124B}, suggesting that the putative neutral hydrogen exosphere of WASP-29~b is quickly accelerated away from the planet in a strong regime of radiative blow-out.

\subsection{Reconstruction of the X-rays and extreme ultraviolet spectrum}

\begin{figure}
\centering
\includegraphics[width=\hsize]{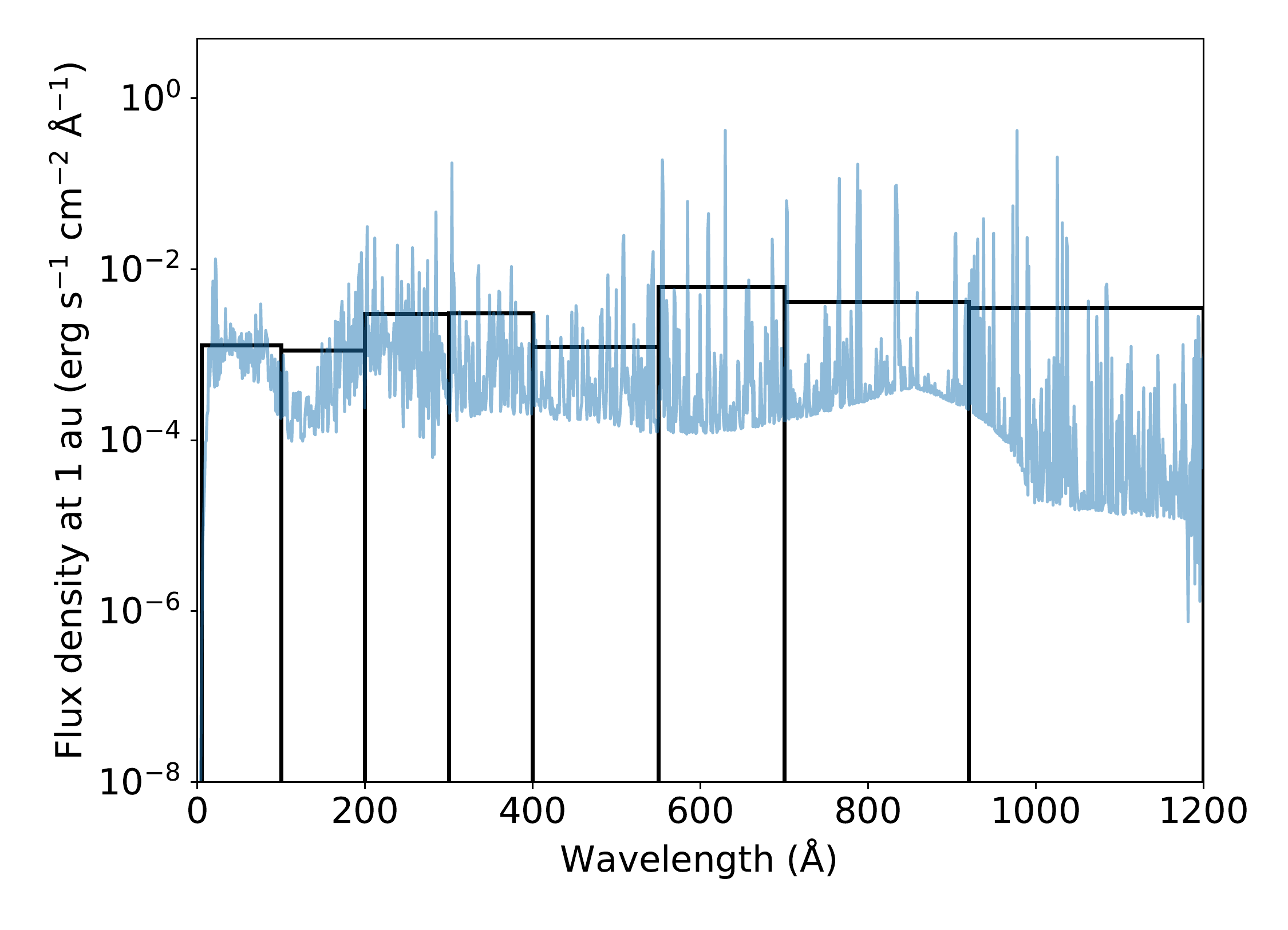} 
\caption{Synthetic X+EUV spectrum of WASP-29 in high resolution (blue) and low resolution (black).}
\label{XEUV_spec}
\end{figure}

The results for COS observations allow us to provide constraints on the X-rays and extreme ultraviolet (X+EUV) emission of WASP-29, a part of its high-energy spectrum that is widely absorbed by the ISM and not straightforward to measure directly. WASP-29 was also observed by \emph{XMM-Newton} (Prop. ID: 80479, PI J. Sanz-Forcada) in 14 May 2017. During the 6.9~ks exposure only an upper-limit detection was made ($S/N=2$) in the European Photon Imaging Camera (EPIC). We measured an X-ray luminosity of $L_{\rm X} < 3.4 \times 10^{26}$~erg\,s$^{-1}$ in the 0.12--2.48~keV spectral range (5--100~\AA). Based on this upper limit, we used a coronal model to construct the synthetic spectrum in the region 1-1200~\AA , following \citet{2011A&A...532A...6S}. We used the EPIC spectra to model the coronal temperatures, and the well-constrained COS fluxes (Sect. \ref{cos_analysis}) to model the transition region. The model-derived fluxes at 1~au in different EUV bands are $2.74$~erg~s$^{-1}$~cm$^{-2}$ (100--920 \AA), and $0.78$~erg~s$^{-1}$~cm$^{-2}$ (100--504~\AA), for the first ionization edges of H and He, respectively (Fig. \ref{XEUV_spec}). The resulting X+EUV (5--920 \AA) flux at 1~au is 3.98~erg~s$^{-1}$~cm$^{-2}$. Further details of the coronal model will be given elsewhere (Sanz-Forcada et al., in prep.). The observed X-ray emission is rather low for a K4V star \citep{piz03}, which implies that WASP-29 is an aged star with a low state of activity.

Using the energy-limited approximation \citep{2016A&A...585L...2S}, we estimate a total mass loss rate of WASP-29~b at $4 \times 10^{9}$~g~s$^{-1}$ for 15\% escape efficiency.  This value is intermediate when compared to the rates inferred from observations of the hot Jupiters HD~209458~b and HD~189733~b, and the warm Neptunes GJ~436~b and GJ~3470~b, which range from $10^8$ to $10^{11}$~g~s$^{-1}$ \citep{2010A&A...514A..72L, 2013A&A...557A.124B, 2016A&A...591A.121B, 2018A&A...620A.147B}.

Similarly to \citet{2017A&A...597A..26B}, we estimate the photoionization rate and lifetime of atomic H around WASP-29~b based on the synthetic spectrum up to the ionization threshold of 911.8~\AA\ (13.6~eV). We found that the \ion{H}{I} photoionization rate is $11.8 \times 10^{-5}$~s$^{-1}$, yielding a photoionization lifetime of 2.35~h. For comparison, the \ion{H}{I} lifetimes around the warm Neptunes with detected exospheres GJ~436~b and GJ~3470~b are $\sim$12~h and 55 minutes, respectively \citep{2016A&A...591A.121B, 2018A&A...620A.147B}; in the case of the hot Jupiters with detected exospheres HD~209458~b and HD~189733~b, the photoionization lifetimes are $\sim$8~h and $\sim$3.5~h, respectively \citep{2013A&A...557A.124B}. The most likely explanation for the non-detection of exospheric \ion{H}{I} in WASP-29~b is a combination of the short photoionization lifetime, the fast acceleration of H atoms away from the planet by radiation pressure (also known as a "blow-out" regime), the low signal-to-noise ratio in the \lya\ blue wing and the strong ISM absorption precluding access to velocities between [-84, +35]~km~s$^{-1}$. Thus, it is possible that WASP-29~b is, in fact, currently losing its atmosphere, but the high-energy environment around the planet and the distance of the system in relation to Earth are not conducive to produce detectable signals in FUV transmission spectroscopy.

One possible alternative to search for atmospheric escape signals in WASP-29~b is through the metastable \ion{He}{I} triplet near 1.083~$\mu$m \citep{2000ApJ...537..916S, 2018ApJ...855L..11O} in transmission. This technique has been successfully applied in both space- and ground-based spectroscopy \citep[e.g.,][]{2018Natur.557...68S, 2018Sci...362.1384A, 2018ApJ...868L..34M, 2018Sci...362.1388N, 2019A&A...629A.110A, 2020AJ....159..115K}. We computed the \ion{He}{I} ionizing irradiation (up to 504~\AA) level of WASP-29~b using the synthetic spectrum, and found a value of 0.44~W~m$^{-2}$ at the planet's orbit ($440$~erg~s$^{-1}$~cm$^{-2}$). For comparison, the hot Jupiter HD~209458~b receives an irradiation of approximately 1~W~m$^{-2}$, and its \ion{He}{I} signal was challenging to measure \citep{2019A&A...629A.110A}; it is possible that a significant amount of observational effort may be required to measure \ion{He}{I} in the extended atmosphere of WASP-29~b, since the intensity of the planetary signal in this line depends on the level of ionizing XUV irradiation received \citep{2018Sci...362.1388N}. The planets GJ~436~b, KELT-9~b \citep{2018Sci...362.1388N}, and WASP-127~b \citep{2020A&A...640A..29D}, which possess lower EUV irradiation levels than WASP-29~b, exhibit non-detections.

\section{Conclusions}\label{conclusions}

We observed six transits of WASP-29~b using the STIS and COS spectrographs installed on \hst\ to search for signals of atmospheric escape in transmission. We did not find evidence for the presence of a \ion{H}{I} exosphere around WASP-29~b, likely due to a combination of factors: short \ion{H}{I} photoionization lifetime, a fast acceleration of H atoms away from the planet, and the low signal-to-noise ratio in the \lya\ emission between Doppler velocities [-100, +100]~km~s$^{-1}$, which is the range where we expect such signal to appear.

An analysis of the stellar flux time series from COS spectra did not reveal significant, repeatable evidence for escaping metallic species in the atmosphere of WASP-29~b. The stellar line fluxes remain mostly stable over the three observing epochs. We found only a tentative in-transit absorption in the ground-state \ion{C}{II} emission near 1334.5 \AA, but we were unable to disentangle it from stellar activity effects. This result reinforces the requirement of repeatability for FUV atmospheric signals since this in-transit absorption is dominated by only one of the three visits. One way to assess if the observed signal comes from planetary absorption is to search for atmospheric escape using the 1.083~$\mu$m metastable He triplet. If He is detected at high-altitudes around WASP-29~b, it is likely the planet is undergoing active atmospheric escape and thus would give support to the hypothesis of an atmospheric \ion{C}{II} signal.

Despite these non-detections, our observations allowed us to constrain the high-energy spectrum of WASP-29, as well as the environment around WASP-29~b. We reconstructed the intrinsic \lya\ spectrum of WASP-29 using the STIS observations, and found that the \lya\ flux is stable when we compare the three different epochs of observation spanning a baseline of 12 days. Furthermore, this results suggests that the level of \lya\ radiation pressure efficiently overcomes gravity by a factor of seven, which is higher than those observed for other G- and K-type exoplanet hosts, such as HD~189733, HD~97658, and HD~209458. This puts the planet WASP-29~b in a regime of faster radiative acceleration of H atoms compared to other hot giant exoplanets for which we have observed signals of \ion{H}{I} escape. Future observations of \ion{He}{I} escape in the 1.083 $\mu$m triplet may be able to provide better atmospheric escape constraints for WASP-29~b. We also used \emph{XMM-Newton} observations to constrain its EUV spectrum using a coronal model. The low levels of stellar activity in FUV, lack of stellar flares (in contrast to, for example, GJ~436), and the inferred low X-ray luminosity strongly suggest that WASP-29 is an old star. Based on the reconstructed EUV spectrum, we estimate that WASP-29~b is currently losing its atmosphere at a total rate of $4 \times 10^9$~g~s$^{-1}$, assuming a heating efficiency of 15\%. This mass loss rate is intermediate between the rates observed in hot Jupiters and warm Neptunes through \lya\ spectroscopy.

\begin{acknowledgements}
L.d.S. thanks the valuable discussions with P. Wheatley on the results of this manuscript. G.W.H. acknowledges long-term support from NASA, NSF, Tennessee State University, and the State of Tennessee through its Centers of Excellence program. A.L.d.E. acknowledges the CNES for financial support. J.S.F. acknowledges support from the Spanish State Research Agency projects AYA2016-79425-C3-2-P and PID2019-109522GB-C51. We are grateful for the helpful feedback provided by the anonymous referee. This project has received funding from the European Research Council (ERC) under the European Union's Horizon 2020 research and innovation programme (project {\sc Four Aces} grant agreement No 724427; project {\sc Spice Dune} grant agreement No 947634), and it has been carried out in the frame of the National Centre for Competence in Research PlanetS supported by the Swiss National Science Foundation (SNSF). This research is based on observations made with the NASA/ESA Hubble Space Telescope. The data are openly available in the Mikulski Archive for Space Telescopes (MAST), which is maintained by the Space Telescope Science Institute (STScI). STScI is operated by the Association of Universities for Research in Astronomy, Inc. under NASA contract NAS 5-26555. This research made use of the NASA Exoplanet Archive, which is operated by the California Institute of Technology, under contract with the National Aeronautics and Space Administration under the Exoplanet Exploration Program. We used the open source software SciPy \citep{scipy_ref}, Jupyter \citep{Kluyver:2016aa}, Astropy \citep{2013A&A...558A..33A}, Matplotlib \citep{Hunter:2007}, \texttt{batman} \citep{2015PASP..127.1161K}, and \texttt{emcee} \citep{2013PASP..125..306F}.
\end{acknowledgements}

\bibliographystyle{aa}
\bibliography{biblio.bib}

\end{document}